\documentstyle[12pt]{article}

\def\be{\begin{equation}}
\def\ee{\end{equation}}
\def\bea{\begin{eqnarray}}
\def\eea{\end{eqnarray}}

\begin{document}

\title{Classical and quantum N=1 super $W_\infty$-algebras}

\author{L. O. Buffon$^{1}$, D. Dalmazi$^{2}$ and A. Zadra$^{1}$\\
{}\\
${}^{1}${\it Instituto de F\'{\i}sica, Universidade de S\~ao Paulo }\\
{\it CP 66318, 05389-970,  S\~ao Paulo, Brazil }\\
{\it lobuffon@uspif.if.usp.br, azadra@uspif.if.usp.br}\\
{}\\
${}^{2}${\it UNESP, Campus de Guaratinguet\'a }\\
{\it CP 205, Guaratinguet\'a, S\~ao Paulo, Brazil}\\
{\it dalmazi@grt000.uesp.ansp.br}
}

\date{}

\maketitle

\begin{abstract}

We construct higher-spin N=1 super algebras as extensions of the 
super Virasoro algebra containing generators for all spins $s\ge 3/2$. 
We find two distinct classical (Poisson) algebras on the phase super 
space. Our results indicate that only one of them can be consistently 
quantized.

\end{abstract}

\section{Introduction}

The infinite-dimensional Virasoro algebra and its extensions play a
fundamental r\^ole in the study of two-dimensional conformal field 
theories. In particular, the $W_N$-algebras \cite{1} are non-linear 
algebras which contain additional generators, corresponding to fields 
with conformal spins $s$ in the interval $2\le s\le N$. 
In contradistinction, the $W_\infty$-type algebras \cite{2} \cite{3}, 
generated by an infinite set of higher-spin operators with $s\ge 1$ 
or $2$, are linear algebras. They appear in the continuum formulation 
of two-dimensional quantum gravity coupled to $c=1$ matter and 
also in some discrete multi-matrix models which are related to the 
$c=1$ theory \cite{4}-\cite{7.5}. 
Our interest in super $W_{\infty}$ algebras was raised in a recent 
paper \cite{8}, where we studied the Schwinger-Dyson (S-D) equations 
of the N=1 supersymmetric eigenvalue model \cite{9}, which is a 
supersymmetric version of the hermitian one-matrix model written in 
terms of eigenvalues. We found a correspondence between those S-D 
equations and the bosonic sector of an N=1 super $W_\infty$-algebra. 
In this work, we aim to characterize the full N=1 super algebra, 
including bosonic and fermionic operators. In fact, we have noticed 
a lack of explicit formulae in the N=1 case, since the literature 
mostly concerns N=2 and some of its reductions \cite{10}-\cite{13}.

We shall start from a classical realization, that is a Poisson 
algebra on a phase super space, with a pair of commuting and 
anti-commuting partners $(x,\theta)$ and their conjugate momenta, 
$(p,\Pi )$ respectively. The ``quantum" algebras, announced in 
the title, will be constructed by replacing momenta by differential 
operators, $p\to -i\hbar \partial /\partial x$ and 
$\Pi \to -i\hbar \partial /\partial \theta $, and Poisson brackets 
by commutators. The Planck's constant $\hbar $ will be used to 
control the classical limit $(\hbar \to 0)$ in the usual way, 
${1\over i\hbar} [\cdot ,\cdot ] \to \{ \cdot, \cdot \}$. 
The spin $s$ of the generators will be classified \cite{13} according 
to their maximal power in momenta (or derivatives): for the bosonic 
operators, the maximal power is $p^{s-1}$; for the fermionic ones, 
we have $p^{s-1/2}$. The phase space (or differential) realization 
is specially suitable for higher-spin extensions, because the Jacobi 
identity (which is rather cumbersome to check for $W_{\infty}$-algebras) 
is already built in and it can be effectively used to derive several 
brackets, so that the calculations become altogether simpler. 

In section 2 we describe the classical $w_\infty$-algebra, two 
supersymmetric extensions and a geometric interpretation. The 
quantization is presented in section 3 and the corresponding classical 
limit is discussed. Section 4 is dedicated to final comments and 
conclusions.

\section{Classical N=1 super $w_\infty$-algebras}

In the bosonic case, the $w_\infty$-algebra is equivalent to the 
Poisson algebra of smooth area-preserving diffeomorphisms on a 
two-dimensional phase space $(x,p)$. Following refs.\cite{2} \cite{3}, 
we introduce the Poisson brackets:

\be
\label{poisson}
\{ f(x,p),g(x,p) \} = {\partial f\over \partial x}{\partial g\over 
\partial p}
 - {\partial f\over\partial p}{\partial g\over \partial x} \quad .
\ee
The area-preserving transformations, which preserve the 2-form 
$\omega = dx\Lambda dp$, correspond to canonical 
transformations generated by smooth functions $\rho (x,p)$ via Poisson 
brackets, $f \to f + \epsilon\{f,\rho(x,p)\} $. The smooth functions 
$\rho $ can be expanded as $\rho = \sum _{s,n} \rho _{sn} w_n^{(s)}$, 
where we take the basis
\be
\label{wns}
w_n^{(s)} = x^{n+1}p^{s-1} \quad .
\ee
This set of functions generate the classical $w_\infty$-algebra 
\cite{2}

\be
\label{wi1}
\{ w_m^{(r)}, w_n^{(s)} \} = \left[ (s-1)(m+1) - (r-1)(n+1) \right]
w_{m+n}^{(r+s-2)} \quad ,
\ee
which can be seen as a higher-spin extension ($s\ge 2$) of the
$s=2$ Virasoro algebra generated by $w_n^{(2)}=x^{n+1}p$.  
Introducing a Grassmann-odd spin-$3/2$ generator $g_n^{(3/2)}$, 
the Virasoro algebra can be extended to a superconformal 
algebra\footnote{Throughout this paper, we shall consider the 
Neveu-Schwarz sector of the superconformal algebra.}:
\bea
\label{superv}
& &\{ g_m^{(3/2)}, g_n^{(3/2)} \} = 2 w_{m+n+1}^{(2)} \quad 
,\nonumber \\
& &\{ g_m^{(3/2)}, w_n^{(2)} \} = \left[ (m+1) - {1\over 
2}(n+1)\right]
g_{m+n}^{(3/2)} \quad ,\\
& &\{ w_m^{(2)}, w_n^{(2)} \} = (m-n) w_{m+n}^{(2)} \quad .\nonumber
\eea
Assuming the canonical graded Poisson brackets,  
$\{ x,p\} = 1 \, ,\, \{ \theta,\Pi \}_+ = -1\, $,
the most general realization for $g_n^{(3/2)}$ and $w_n^{(2)}$, 
which is compatible with the infinitesimal conformal transformations

\bea
\label{conftr}
& \delta x &= \{x, \epsilon w_n^{(2)} + \alpha g_n^{(3/2)}\} = 
\epsilon x^{n+1} + \alpha \theta x^{n+1}\quad ,\\
& \delta \theta &= \{\theta, \epsilon w_n^{(2)}+ \alpha g_n^{(3/2)}\} = 
\epsilon {(n+1)\over 2} x^{n}\theta + \alpha x^{n+1}\quad ,
\eea
and with the algebra (\ref{superv}), is given by\footnote{Actually, 
the most general form would be
$ g_n^{(3/2)} = x^{n+1}(\theta p-\Pi) + (2\lambda n + \gamma 
)x^n\theta $, 
but the parameter $\gamma $ can be shifted by a canonical automorphism 
generated by $\gamma {\rm ln}|x|$. Therefore, we may take e.g. 
$\gamma = 2\lambda $ with no loss of generality.}:

\bea
\label{g32}
&g_n^{(3/2)}(\lambda ) &= x^{n+1}(\theta p-\Pi ) + 2\lambda 
(n+1)x^n\theta \quad ,
\\
&w_n^{(2)}(\lambda ) &= x^{n+1}p + (n+1)x^n\left( \lambda + {\theta 
\Pi
\over 2}\right) \quad ,
\eea
where $\lambda$ is an arbitrary real constant.

To include higher-spin generators and extend the super Virasoro 
algebra (\ref{superv}), we make the following assumptions:

i) The lowest spin is $s=3/2$.

ii) There exists a fermionic generator with spin $s=5/2$.

iii) The Poisson algebra of fermionic generators must obey the rule:
\[
\{ g^{(r)}, g^{(s)} \} \propto w^{(r+s-1)} + {\rm lower\; spins} 
\quad .
\]

iv) Each generator $g_n^{(s)}$ is characterized by two indices: $s$ 
corresponds to its spin, and $n$ to its conformal dimension (the 
eigenvalue of $L_0=w_0^{(2)}$).

We try the most general Ansatz for the next-spin generator, $g_n^{(5/2)}$, 
in agreement with the assumptions i)-iv), such that the algebra with 
$g_n^{(3/2)}$ gets closed:
\be
g_{m-1}^{(5/2)}\, = \, x^m\theta p^2 + c_m x^m p\Pi + d_m x^{m-1}\Pi + 
e_m x^{m-2}\theta \quad .
\ee
In order to calculate the arbitrary constants $c_m,d_m,e_m$ we
verify that:
\bea
\label{rst}
&\{ g_{n-1}^{(3/2)},g_{m-1}^{(5/2)} \} \, &= \, 
(d_m + 2\lambda nc_m)w_{n+m-2}^{(2)} + (c_m -1)x^{n+m}p^2 + \nonumber \\
&  &+ R_{nm}x^{n+m-1}p\Pi\theta + 
S_{nm} x^{n+m-2} (\theta\Pi - 2\lambda) + T_{nm} x^{n+m-2} \quad , 
\eea
where $R_{nm},S_{nm},T_{nm}$ are given functions of $n,m,c_m,d_m,e_m$.
The next step is to determine the most general linear combinations
of the terms on the r.h.s. of (\ref{rst}) (except, of course,  
$w_{n+m-2}^{(2)}$ which is already in the algebra), so that they close 
the algebra with $g_n^{(3/2)}$. We define such combinations as: 
\be
V_m^{(3)} \, = \, a_m x^mp^2 + f_m x^{m-1}p\Pi\theta + g_m x^{m-
2}(\theta\Pi-2\lambda) + h_mx^{m-2} \quad .
\ee
It is easy to see that $\{ g_{n-1}^{(3/2)},V_m^{(3)} \}$
will produce the term $x^{n+m-1}\theta p^2$, among others with lower spins.
This term must be part of $g_{n+m-2}^{(5/2)}$ due
to the uniqueness of the spin-$5/2$ generator. Indeed, if there were 
more than one solution for $c_m,d_m,e_m$ for a given algebra, either 
the assumption i) or iii) would fail. Therefore, the closure 
of $\{g_{n-1}^{(3/2)}\, ,\, V_m^{(3)}\}$ imposes the following
conditions:
\bea
\label{amfm}
&\left( (m-2n)a_m + f_m\right) c_{m+n-1} &=2na_m + f_m \quad ,\\
&\left( (m-2n)a_m + f_m\right) d_{m+n-1}
&=2\left( \lambda n (f_m - 2(n-1)a_m)- g_m\right) \quad ,
 \\
&\left( (m-2n)a_m + f_m\right) e_{m+n-1}
&=(m-2)h_m + 4n\lambda^2(n+m-2)(2(n-1)a_m - f_m) .
\eea
These equations require that $\lambda =0$ and
we find two possible Ansatze for $g_n^{(5/2)}$, corresponding
to two different algebras. All higher spins
generators, $w_{n}^{(s)}$ $(s\ge 2)$ and $g_n^{(s)}$ $(s>5/2)$,
are obtained from $s\le 5/2$ generators. We end up
with $\lambda =0$ and two possible N=1 supersymmetric $w$-algebras:
\vskip .3cm

{\bf Type 1:} This type is generated by
\be
\label{chg}
g_n^{(k+3/2)} = x^{n+1}p^k(\theta p -\Pi)\quad ,
\ee
\be
\label{chw}
w_n^{(s)} = x^{n+1}p^{s-1} + {1\over 2}(n+1)x^np^{s-2}\theta \Pi 
\quad ,
\ee
with the following algebra:
\bea
\label{superch}
& &\{ g_m^{(r)}, g_n^{(r')} \} = 2 w_{m+n+1}^{(r+r'-1)} \quad 
,\nonumber \\
& &\{ g_m^{(r)}, w_n^{(s)} \} = \left[ (s-1)(m+1) - (r-1)(n+1)\right]
g_{m+n}^{(r+s-2)} \quad ,\\
& &\{ w_m^{(s)}, w_n^{(s')} \} = \left[ (s'-1)(m+1) - (s-
1)(n+1)\right]
w_{m+n}^{(s+s'-2)} \quad .\nonumber
\eea
where $r,r'=3/2,5/2,\cdots $ and $s,s'=2,3,4,\cdots$. In fact, 
this algebra appeared in \cite{11}, in a more complicated realization.

\vskip .5cm
{\bf Type 2:} In this case, the generators split in four families 
with only even spins in the bosonic sector. The generators are given 
by:
\be
\label{syg}
g_n^{(2a+3/2)} = x^{n+1}p^{2a}(\theta p-\Pi )\quad ,
\ee
\be
\label{sygb}
{\overline g}_n^{((2a+1)+ 3/2)} = x^{n+1}p^{2a+1}(\theta p+\Pi )\quad ,
\ee
\be
\label{syw}
w_n^{(2a+2)} = x^{n+1}p^{2a+1} + {1\over 2}(n+1)x^np^{2a}\theta \Pi 
\quad,
\ee
\be
\label{syk}
k_n^{(2a+2)} = x^{n+1}p^{2a+1}\theta \Pi \quad ,
\ee
with the corresponding classical algebra,
\bea
\label{supersy}
& &\{ g_m^{(r)}, g_n^{(r')} \} = 2 w_{m+n+1}^{(r+r'-1)} 
\quad ,
\nonumber \\
& &\{ g_m^{(r)}, w_n^{(s)} \} = \left[ (s-1)(m+1) - 
(r-1)
(n+1)\right] g_{m+n}^{(r+s-2)} \quad ,\nonumber\\
& &\{ w_m^{(s)}, w_n^{(s')} \} = \left[ (s'-1)(m+1) - 
(s-1)(n+1)\right]
 w_{m+n}^{(s+s'-2)} \quad ,\nonumber \\
& &\{ {\overline g}_m^{(r)}, {\overline g}_n^{(r')} \} = -2
w_{m+n+1}^{(r+r'-1)} \quad ,\nonumber \\
& &\{ {\overline g}_m^{(r)}, w_n^{(s)} \} = \left[ 
(s-1)(m+1) -
(r-1)(n+1)\right] {\overline g}_{m+n}^{(r+s-2)} \quad ,
\nonumber\\
& &\{ g_m^{(r)}, {\overline g}_n^{(r')} \} = 2\left[ 
(r'-1)
(m+1) - (r-1)(n+1)\right] k_{m+n}^{(r+r'-2)} \quad 
,\nonumber \\
& &\{ g_m^{(r)}, k_n^{(s)}\} = {\overline 
g}_{m+n+1}^{(r+s-1)}
\quad ,\nonumber \\
& &\{ {\overline g}_m^{(r)}, k_n^{(s)}\} = 
g_{m+n+1}^{(r+s-1)}
\quad ,\nonumber \\
& &\{ k_m^{(s)}, k_n^{(s')} \} = 0 \quad ,\nonumber \\
& &\{ k_m^{(s)}, w_n^{(s')} \} = \left[ (s'-1)(m+1) - 
(s-1)(n+1)\right]
 k_{m+n}^{(s+s'-2)} \quad .
\eea
As far as we know, this algebra has not appeared yet in the
literature and we shall call it {\it super even} 
$w_\infty${\it -algebra}.
We note, in passing, that the two algebras (\ref{superch}) and 
(\ref{supersy}) have a sub-algebra in common, generated by 
$g_n^{(2a+3/2)}$ and $w_m^{(2a)}$. This algebra is called super 
$w_{\infty \over 2}$, since its bosonic sector corresponds to the 
$w_{\infty}$ truncated to even spins, i.e. $w_{\infty\over 2}$.

Both Poisson algebras are related to area-preserving diffeomorphisms:
they preserve the 2-form $w=dx\Lambda dp - d\Pi\Lambda d\theta $ 
\cite{10}. The super $w_{\infty}$-algebra corresponds to 
transformations generated by the following kind of functions:
\be
\label{rhoch}
\rho_{A} = \phi (x + {\theta \Pi\over 2p}, p) + (\theta p-\Pi )\psi
(x,p)
\quad ,
\ee
while the super even algebra is related to generating functions of 
the
form:
\be
\label{rhosy}
\rho_{B} = p\phi (x + {\theta \Pi \over 2p}, p^2) + \theta \Pi 
p\varphi
(x,p^2) + (\theta p-\Pi )\psi (x,p^2) + (\theta p+\Pi) p\eta (x,p^2)
\quad .
\ee
Above, $\phi, \varphi, \psi$ and $\eta$ are smooth functions of two
variables. These generators correspond to two different invariant 
sub-groups of (super)area-preserving diffeomorphisms. 
In fact, if $\rho_1$ and $\rho_2$ have the form (\ref{rhoch}), so 
will have $\rho_3 = \{ \rho_1,\rho_2 \}$. An analogous result holds 
for functions of the type (\ref{rhosy}). We recall that, 
in a general basis, for arbitrary smooth functions 
$\rho (x,p, \theta, \Pi)$, one finds an  
N=2 super $w_{\infty}$-algebra (see \cite{11} \cite{14}).

\section{Quantum N=1 super $W_\infty$-algebra}

By ``quantum" algebra we mean algebra of commutators, as a quantized
version of the Poisson algebras analyzed in section 2. In the bosonic 
case, the Virasoro algebra is generated by the differential operators
\be
\label{W2}
L_n \equiv W_n^{(2)} = -i\hbar x^{n+1}\partial \quad ,
\ee
obtained from its classical counterpart $w_n^{(2)}$ in (\ref{wns}) 
after the replacement $p \to -i\hbar \partial $. The set of higher spin 
operators
\be
\label{Wns}
W_n^{(s)} = (-i\hbar )^{s-1} x^{n+1}\partial ^{s-1} \quad , \quad 
s\ge 1
\quad ,
\ee
generate the so called $W_{1+\infty}$-algebra, given by:
\be
\label{W1pi}
[W_m^{(r)}, W_n^{(s)}]= -i\hbar \sum _{k\ge 0} (-i\hbar )^k 
C_{mn}^{rs}(k)
W_{m+n-k}^{(r+s-2-k)} \quad ,
\ee
\be
\label{C}
C_{mn}^{rs}(k) = {1\over (k+1)!}\left(
{\Gamma(r) \over \Gamma(r-k-1)}{\Gamma(n+2)\over \Gamma(n-k+1)} -
{\Gamma(s) \over \Gamma(s-k-1)}{\Gamma(m+2)\over \Gamma(m-k+1)}
\right) \quad .
\ee
The generator (\ref{W2}) can be generalized\footnote{The operators 
$L_n(\lambda ) = -i\hbar (x^{n+1}\partial + (a+\lambda n)x^n)$
also generate the Virasoro algebra. However, the parameter $a$ can be
arbitrarily shifted by the homeomorphism $L_n\to$  $x^{i\hbar c}L_n x^
{-i\hbar c}$  $\Longrightarrow$ $a\to a+c$. Thus, we may take 
$a=\lambda$.} into the form (see \cite{12})
\be
\label{cw2l}
W_n^{(2)}(\lambda ) = -i\hbar \left( x^{n+1}\partial + \lambda (n+1)
x^n\right) \quad .
\ee
We are interested in a basis of operators which satisfy the following 
condition (originally used to discover the $W_{\infty}$-algebra [3]):
\be
\label{parcon}
[W_m^{(r)},W_n^{(s)}] = -i\hbar \left( c_oW_{m+n}^{(r+s-2)} + 
c_1W_{m+n-2}^
{(r+s-4)} + \cdots \right) \quad .
\ee
This sort of basis is convenient because the algebra can be truncated 
in only even-spin sub-algebras. Moreover, it admits a central extension 
\cite{3}.

The condition (\ref{parcon}) restricts the possible values of the
parameter $\lambda $. In analogy to the last section, we take 
an Ansatz for $W_m^{(3)}$ and we find two solutions (in agreement 
with \cite{14}):
\vskip .3cm

i) If $s\ge 1$, we have $\lambda = 1/2$ and the $W_{1+\infty}$-algebra. 
The first few generators are given below:
\bea
\label{W1piex}
& &W_n^{(1)} = x^{n+1}\quad ,\nonumber \\
& &W_n^{(2)} = (-i\hbar) \left( x^{n+1}\partial + {1\over 2}(n+1)x^n
\right) \quad ,\nonumber \\
& &W_n^{(3)} = (-i\hbar)^2 \left( x^{n+1}\partial ^2 + 
(n+1)x^n\partial
\right) \quad ,\nonumber \\
& &W_n^{(4)} = (-i\hbar)^3 \left( x^{n+1}\partial ^3 + {3\over 2}(n+1)
x^n\partial ^2 +{1\over 2}n(n+1)x^{n-1}\partial \right) \quad .
\eea
Higher spin operators can be obtained via commutators.

\vskip .3cm
ii) If $s\ge 2$, one has two equivalent cases, $\lambda =0$ or $1$.
When $\lambda =0$ one finds a $W_\infty$-algebra, generated by:
\bea
\label{Wi}
& &W_n^{(2)} = (-i\hbar)\; x^{n+1}\partial \quad ,\nonumber \\
& &W_n^{(3)} = (-i\hbar)^2 \left( x^{n+1}\partial ^2 + {1\over 2}(n+1)
x^n\partial \right) \quad ,\nonumber \\
& &W_n^{(4)} = (-i\hbar)^3 \left( x^{n+1}\partial ^3 + 
(n+1)x^n\partial
^2 \right) \quad ,\quad {\rm etc. } 
\eea
The solution $\lambda =1$ corresponds to an automorphism of the above
generators, leading to an isomorphic $W_\infty$-algebra.
\vskip .3cm

Now we present the N=1 supersymmetric extension of the $W_\infty$-
algebra. First, we introduce an anti-commuting variable $\theta $ 
and proceed in analogy to the classical study, by making the 
following assumptions:

i) The lowest spin $(s=3/2)$ generator \cite{12} \cite{13} is
\be
\label{G32l}
G_n^{(3/2)} = (-i\hbar) \left( x^{n+1}(\theta \partial - \partial 
_\theta
) + 2\lambda (n+1)x^n\theta \right) \quad .
\ee
Together with the spin-2 operator,
\be
\label{ccw2l}
W_n^{(2)} = (-i\hbar) \left( x^{n+1}\partial + {1\over 
2}(n+1)x^n(\theta
\partial _\theta + 2\lambda )\right) \quad ,
\ee
they generate the super Virasoro algebra:
\bea
\label{sva}
& &[G_m^{(3/2)}, G_n^{(3/2)}] = 2i\hbar W_{m+n+1}^{(2)} \quad 
,\nonumber \\
& &[G_m^{(3/2)}, W_n^{(2)}]= i\hbar \left( (m+1) - {1\over 2}(n+1) 
\right)
G_{m+n}^{(3/2)} \quad , \\
& &[W_m^{(2)}, W_n^{(2)}] = i\hbar (m-n) W_{m+n}^{(2)} \quad 
,\nonumber
\eea
whose classical limit coincides with the algebra (\ref{superv}).

\vskip .3cm
ii) We assume the existence of a spin-5/2 generator, whose most 
general
expression is:
\be
\label{G52}
G_n^{(5/2)} = (-i\hbar)^2 \left( x^{n+1}\partial (\theta \partial + 
c_n
\partial _\theta ) + d_n x^n\partial _\theta + e_n x^{n-1}\theta 
\right)
\quad ,
\ee
where the constants $c_n, d_n, e_n$ must be determined.

\vskip .3cm
iii) The anti-commutation algebra should obey the rule:
\be
\label{spc}
[G_m^{(r)}, G_n^{(s)}] \propto W_{m+n+1}^{(r+s-1)} + {\rm lower\; 
spins}
\quad .
\ee

\vskip .3cm
iv) Each operator $G_n^{(s)}$ is characterized by its spin $(s)$ and 
its
conformal dimension $(n)$.

\vskip .3cm
Under these assumptions, we find two solutions, $\lambda =0$ or $1/2$,
which are related to each other by an automorphism. Therefore, we may
simply take $\lambda = 0$ and the resulting N=1 super 
$W_\infty $-algebra can be generated by the following basis of 
operators (we present the lowest spins, since higher spins can be
produced by commutators):
\bea
\label{swi1pi}
& &G_n^{(3/2)} = (-i\hbar ) x^{n+1}(\theta \partial -\partial _\theta 
)
\quad ,\nonumber \\
& &G_n^{(5/2)} = (-i\hbar )^2 \left( x^{n+1}\partial (\theta \partial 
+
\partial _\theta ) + (n+1)x^n\partial _\theta \right) \quad 
,\nonumber \\
& &G_n^{(7/2)} = (-i\hbar )^3 \left( x^{n+1}\partial ^2(\theta 
\partial
- \partial _\theta ) - 2(n+1)x^n\partial \partial _\theta -n(n+1)x^{n-
1
}\partial _\theta \right) \quad ,\nonumber \\
& &G_n^{(9/2)} = (-i\hbar )^4 \left( x^{n+1}\partial ^3(\theta 
\partial
+ \partial _\theta ) + 3(n+1)x^n\theta \partial ^3 +3n(n+1)x^{n-
1}\theta
\partial ^2\right.\nonumber \\
& &\left. \qquad  + (n-1)n(n+1)x^{n-2}\partial _\theta \right) \quad ,
\nonumber \\
& &W_n^{(2)} = (-i\hbar )\left( x^{n+1}\partial +{1\over 
2}(n+1)x^n\theta
\partial _\theta \right) \quad ,\nonumber \\
& &K_n^{(2)} = (-i\hbar )^2 x^{n+1}\partial \partial _\theta \theta
\quad ,\nonumber \\
& &W_n^{(4)} = (-i\hbar )^3 \left( x^{n+1}\partial ^3 +{3\over 2}(n+1)
x^n \partial ^2 +{1\over 2}n(n+1) x^{n-1}\partial \right. \nonumber \\
& &\left. \qquad -{1\over 2}(n+1)x^n \partial ^2
\partial _\theta \theta \right) \quad ,\nonumber \\
& &K_n^{(4)} = (-i\hbar )^4 \left( x^{n+1}\partial^3 + (n+1) x^n
\partial ^2\right) \partial _\theta \theta \quad ,\nonumber \\
& &W_n^{(6)} = (-i\hbar )^5 \left( x^{n+1}\partial ^5 +{5\over 2}(n+1)
x^n \partial ^4 + 2n(n+1) x^{n-1}\partial ^3 \right. \nonumber \\
& & \left. \qquad + {1\over 2}(n-1)n(n+1)x^{n-2}\partial ^2(1 +
\partial _\theta \theta ) -{1\over 2}(n+1)x^n\partial ^4 \partial 
_\theta
\theta \right) \quad ,\nonumber \\
& &K_n^{(6)} = (-i\hbar )^6 \left( x^{n+1}\partial^5 + 2(n+1) x^n
\partial ^4 + n(n+1)x^{n-1}\partial ^3 \right) \partial _\theta 
\theta \quad .
\eea
We calculated various commutators (up to spin $s=6$; further
commutators can be obtained by means of the Jacobi identity), but we
were unable to find a closed form for all structure coefficients.
The lowest-spin algebra is listed below:
\bea
\label{algebra}
& &[G_m^{(3/2)}, G_n^{(3/2)}] = 2i\hbar W_{m+n+1}^{(2)} \quad 
,\nonumber \\
& &[G_m^{(3/2)}, W_n^{(2)}]= i\hbar ( (m+1) - {1\over 2}(n+1) )
G_{m+n}^{(3/2)} \quad , \nonumber \\
& &[W_m^{(2)}, W_n^{(2)}] = i\hbar (m-n) W_{m+n}^{(2)} \quad 
,\nonumber \\
& &[G_m^{(5/2)}, G_n^{(3/2)}] = i\hbar (m-3n-2)K_{m+n}^{(2)} + 2(-
i\hbar )^2
(m-n)W_{m+n}^{(2)}\quad ,\nonumber \\
& &[G_m^{(3/2)}, K_n^{(2)}] = -i\hbar G_{m+n+1}^{(5/2)}
+ (-i\hbar)^2 (n+1)G_{m+n}^{(3/2)} \quad ,\nonumber \\
& &[K_m^{(2)}, K_n^{(2)}] = (-i\hbar)^2 (n-m)K_{m+n}^{(2)} \quad ,
\nonumber \\
& &[W_m^{(2)},K_n^{(2)}] = i\hbar (m-n)K_{m+n}^{(2)} \quad ,\nonumber 
\\
& &[G_m^{(5/2)}, W_n^{(2)}] = i\hbar (m-{3\over 2}n - {1\over 2}) 
G_{m+n}
^{(5/2)} + (-i\hbar )^2 n(n+1) G_{m+n-1}^{(3/2)} \quad ,\nonumber \\
& &[G_m^{(5/2)}, K_n^{(2)}] = -i\hbar G_{m+n+1}^{(7/2)} + 2(-i\hbar 
)^2
(n+1)G_{m+n}^{(5/2)} + (-i\hbar )^3 n(n+1)G_{m+n-1}^{(3/2)} \quad,
\nonumber \\
& &[G_m^{(5/2)},G_n^{(5/2)}] = -2i\hbar W_{m+n+1}^{(4)}
-3(-i\hbar )^2(n(n+1)+m(m+1))K_{m+n-1}^{(2)} \nonumber \\
& & \qquad +2(-i\hbar )^3 ((n+m+1)(n+m+2) - 3(n+1)(m+1))W_{m+n-
1}^{(2)}
\quad ,\nonumber \\
& &[G_m^{(3/2)},W_n^{(4)}] = -i\hbar {1\over 2}
(n-6m-5)G_{m+n}^{(7/2)} \nonumber \\
& &\qquad +(-i\hbar )^2 (n(n+1) - 3(m+1)(m+n+1))G_{m+n-1}^{(5/2)}
\nonumber \\
& &\qquad +(-i\hbar )^3 {1\over 2}
(m(m+1)(3n+2m+1) + n(n+1)(n-m-2)) G_{m+n-2}^{(3/2)}
\quad .
\eea

We have also verified that the operators in (\ref{swi1pi}) become the
generators (\ref{syg}-\ref{syk}) in the classical limit
(given by the associations $-i\hbar \partial \to p$, $-i\hbar
\partial _\theta \to \Pi$, when $\hbar \to 0$). Therefore, we may say
that the generators (\ref{swi1pi}) realize a (quantum) N=1 super
$W_\infty$-algebra.

Concerning the bosonic sector, composed by $W_n^{(2s)}$ and 
$K_m^{(2r)}$, it is possible to take linear combinations and find a 
basis with two decoupled sub-algebras \cite{8}. For instance, if we 
define
\be
\label{Wt2}
\widetilde W_n^{(2)} = K_n^{(2)} + i\hbar W_n^{(2)}  \quad ,
\ee
the resulting lowest-spin algebra becomes
\bea
\label{VtV}
& &[\widetilde W_m^{(2)}, \widetilde W_n^{(2)}] = (-i\hbar )^2 (n-m)
\widetilde W_{m+n}^{(2)}
 \quad ,\nonumber \\
& &[\widetilde W_m^{(2)}, K_n^{(2)}] = 0 \quad , \\
& &[K_m^{(2)}, K_n^{(2)}] = (-i\hbar )^2 (n-m) K_{m+n}^{(2)} \quad .
\nonumber
\eea
This decoupling was also verified for higher spins. The redefined
$\widetilde W$-operators turn out to generate an algebra isomorphic 
to the even-spin sector of the bosonic $W_{1+\infty}$-algebra. On the 
other hand, the algebra of the operators $K_n^{(2r)}$ is 
isomorphic to the even-spin subalgebra of the 
$W_{\infty}$-algebra\footnote{In the quantum case, 
we may choose a unit system where $\hbar =1$.}. 
Therefore, the bosonic sector of the super algebra 
generated by (\ref{swi1pi}) realizes a $\left( W_{\infty \over 2} 
\oplus W_{{1+\infty} \over 2} \right)$-algebra \cite{8} \cite{12} 
\cite{13}.

It is tempting to call ``N=1 super $\left( W_{\infty \over 2} \oplus
W_{{1+\infty}\over 2} \right)$-algebra" the one generated by the whole
set of operators in (\ref{swi1pi}). We believe this is acceptable at
the quantum level, i.e. as long as $\hbar \not =0$. However, in the 
classical limit $(\hbar \to 0)$ the transformation (\ref{Wt2}) does 
not give independent generators and the bosonic sector does not split 
in two decoupled sub-algebras. This implies that the Poisson algebra 
generated by (\ref{syg}-\ref{syk}) should not be called a ``classical 
super $\left( w_{\infty \over 2} \oplus w_{{1+\infty}\over 2} \right)$" 
-- we had better keep the name N=1 super even $w_\infty$-algebra.

\section{Final remarks, conclusion and open questions}

We have constructed the N=1 supersymmetric extensions of the
$W_\infty$-algebras. At the classical level, we found two Poisson
algebras, the super $w_\infty$ and the super even $w_\infty$. 
In the quantum case, we found only one consistent algebra, denominated 
super $\left( W_{\infty \over 2} \oplus W_{{1+\infty} 
\over 2} \right)$. Its classical limit coincides with the 
super even $w_\infty$. The algebra $\left( W_{\infty \over 2} 
\oplus W_{{1+\infty} \over 2} \right)$ was first observed in \cite{11} as 
a truncation of a N=2 super $W_{\infty}$-algebra. We stress that we 
obtained it in a constructive way, without any embedding 
in higher algebras.

Although we did not find a general expression for all the quantum
operators, we noticed that the available generators can be rewritten
in reduced forms. For instance, the bosonic $K$-operators in
(\ref{swi1pi}) can be expressed as
\be
\label{Ko}
K_n^{(2r)} = (-i\hbar )^{2r} \partial ^{r-1} x^{n+1} \partial ^r
\partial _\theta \theta = p^{r-1}x^{n+1}p^r\Pi\theta \quad .
\ee
The fermionic operators in (\ref{swi1pi}) can be written as linear
combinations of
\be
\label{Go}
\widetilde G_n^{(s+1/2)} = (-i\hbar)^s\left( \partial ^{s-
1}x^{n+1}\theta
\partial + (-)^s x^{n+1} \partial ^{s-1}\partial _\theta \right) =
p^{s-1}x^{n+1}\theta p + (-)^sx^{n+1}p^{s-1}\Pi \quad .
\ee
Therefore, we expect the quantum operators to correspond to some 
special ordering of the classical generators. If we could understand 
this ordering we might eventually find a closed form for the complete 
algebra.

We have shown how the super even $w_\infty$-algebra can be 
obtained from the quantum  super $\left( W_{\infty \over 2} \oplus 
W_{{1+\infty} \over 2} \right)$ by means of a suitable limit 
$(\hbar \to 0)$. It is natural to ask whether there is any quantum 
super $W_\infty$-algebra whose classical limit is the super 
$w_\infty$ given by eqs.(\ref{chg}-\ref{superch}). We do not have an
answer to that question yet. It would also be interesting to study the
possible central extensions \cite{3} of these quantum algebras.

\vfill \eject
\noindent {\bf Acknowledgements}
\vskip 1truecm

The work of L.O.B. and A.Z. and part of the work of D.D. were 
supported by CNPq and Fapesp.

\end{document}